\documentclass{article}
\usepackage{array}

\usepackage{amsmath,amssymb,natbib}

\newcommand{\BEAS}{\begin{eqnarray*}}
\newcommand{\EEAS}{\end{eqnarray*}}
\newcommand{\BEA}{\begin{eqnarray}}
\newcommand{\EEA}{\end{eqnarray}}
\newcommand{\BEQ}{\begin{equation}}
\newcommand{\EEQ}{\end{equation}}
\newcommand{\BIT}{\begin{itemize}}
\newcommand{\EIT}{\end{itemize}}
\newcommand{\BNUM}{\begin{enumerate}}
\newcommand{\ENUM}{\end{enumerate}}

\newcommand{\BA}{\begin{array}}
\newcommand{\EA}{\end{array}}
\newcommand{\BC}{\begin{center}}
\newcommand{\EC}{\end{center}}

\newcommand{\ie}{{\it i.e.}}

\newcommand{\symm}{{\mbox{\bf S}}}  

\newcommand{\Rank}{\mathop{\bf Rank}}
\newcommand{\Tr}{\mathop{\bf Tr}}

\newcommand{\Expect}{{\bf E}}

\usepackage{dsfont}

\def\bx{\mathbf{x}}

\usepackage{enumitem}
\usepackage{enumerate}
\usepackage{amsmath}

\def\RR{\mathbb{R}}

\def\NN{\mathds{N}}

\def\Acal{\mathcal{A}}

\def\one{\mathds{1}}

\def\OMIT#1{}

\DeclareMathOperator{\defi}{def}

\DeclareMathOperator{\defeq}{\overset{\defi}{=}}

\usepackage{enumerate,multirow}

\providecommand{\abs}[1]{\lvert#1\rvert} 
\providecommand{\norm}[1]{\lVert#1\rVert}

\usepackage{tabularx}

\makeatletter
\newif\if@borderstar
\def\bordermatrix{\@ifnextchar*{%
  \@borderstartrue\@bordermatrix@i}{\@borderstarfalse\@bordermatrix@i*}%
}
\def\@bordermatrix@i*{\@ifnextchar[{%
  \@bordermatrix@ii}{\@bordermatrix@ii[()]}
}
\def\@bordermatrix@ii[#1]#2{%
  \begingroup
    \m@th\@tempdima8.75\p@\setbox\z@\vbox{%
      \def\cr{\crcr\noalign{\kern 2\p@\global\let\cr\endline }}%
      \ialign {$##$\hfil\kern 2\p@\kern\@tempdima & \thinspace %
      \hfil $##$\hfil && \quad\hfil $##$\hfil\crcr\omit\strut %
      \hfil\crcr\noalign{\kern -\baselineskip}#2\crcr\omit %
      \strut\cr}}%
    \setbox\tw@\vbox{\unvcopy\z@\global\setbox\@ne\lastbox}%
    \setbox\tw@\hbox{\unhbox\@ne\unskip\global\setbox\@ne\lastbox}%
    \setbox\tw@\hbox{%
      $\kern\wd\@ne\kern -\@tempdima\left\@firstoftwo#1%
        \if@borderstar\kern2pt\else\kern -\wd\@ne\fi%
      \global\setbox\@ne\vbox{\box\@ne\if@borderstar\else\kern 2\p@\fi}%
      \vcenter{\if@borderstar\else\kern -\ht\@ne\fi%
        \unvbox\z@\kern-\if@borderstar2\fi\baselineskip}%
        \if@borderstar\kern-2\@tempdima\kern2\p@\else\,\fi\right\@secondoftwo#1 $%
    }\null \;\vbox{\kern\ht\@ne\box\tw@}%
  \endgroup
}
\makeatother

\usepackage{graphicx}
\usepackage{xcolor}
\definecolor{darkblue}{rgb}{0,0,0.5} 
\usepackage{transparent}
\usepackage{amsfonts,color}
\usepackage{url}
\usepackage{tabularx}
\usepackage{dsfont}
\usepackage{epsf}
\def\Hca{\mathcal{H}}
\def\por{\text{por}}

\begin{document}
\title{Mean-Reverting Portfolios: \\Tradeoffs Between Sparsity and Volatility}%
\author{
\normalsize Marco Cuturi\\
\normalsize Graduate School of Informatics\\
\normalsize Kyoto University\\
\texttt{\small mcuturi@i.kyoto-u.ac.jp}
\\\\
\normalsize Alexandre d'Aspremont\\
\normalsize D.I., UMR CNRS 8548\\
Ecole Normale Sup\'erieure,
\texttt{\small aspremon@ens.fr}
}	
\maketitle
\begin{abstract} Mean-reverting assets are one of the holy grails of financial markets: if such assets existed, they would provide trivially profitable investment strategies for any investor able to trade them, thanks to the knowledge that such assets oscillate predictably around their long term mean. The modus operandi of cointegration-based trading strategies~\citep[\S8]{tsay2005analysis} is to create first a portfolio of assets whose aggregate value mean-reverts, to exploit that knowledge by selling short or buying that portfolio when its value deviates from its long-term mean.  Such portfolios are typically selected using tools from cointegration theory~\citep{granger,johansen}, whose aim is to detect combinations of assets that are stationary, and therefore mean-reverting. We argue in this work that focusing on stationarity only may not suffice to ensure profitability of cointegration-based strategies. While it might be possible to create synthetically, using a large array of financial assets, a portfolio whose aggregate value is stationary and therefore mean-reverting, trading such a large portfolio incurs in practice important trade or borrow costs. Looking for stationary portfolios formed by many assets may also result in portfolios that have a very small volatility and which require significant leverage to be profitable. We study in this work algorithmic approaches that can take mitigate these effects by searching for maximally mean-reverting portfolios which are sufficiently sparse and/or volatile.\end{abstract}
	
\section{Introduction}
	
Mean-reverting assets, namely assets whose price oscillates predictably around a long term mean, provide investors with an ideal investment opportunity. Because of their tendency to pull back to a given price level, a naive contrarian strategy of buying the asset when its price lies below that mean, or selling short the asset when it lies above that mean can be profitable. Unsurprisingly, assets that exhibit significant mean-reversion are very hard to find in efficient markets. Whenever mean-reversion is observed in a single asset, it is almost always impossible to profit from it: the asset may typically have very low volatility, be illiquid, hard to short-sell, or its mean-reversion may occur at a time-scale (months, years) for which the borrow-cost of holding or shorting the asset may well exceed any profit expected from such a contrarian strategy.

\subsubsection{Synthetic Mean-Reverting Baskets} Since mean-reverting assets rarely appear in liquid markets, investors have focused instead on creating synthetic assets that can mimic the properties of a single mean-reverting asset, and trading such synthetic assets as if they were a single asset. Such a synthetic asset is typically designed by combining long and short positions in various liquid assets to form a \emph{mean-reverting portfolio}, whose aggregate value exhibits significant mean-reversion. 

Constructing such synthetic portfolios is, however, challenging. Whereas simple descriptive statistics and unit-root test procedures can be used to test whether a single asset is mean-reverting, building mean-reverting portfolios requires finding a proper vector of algebraic weights (long and short positions) that describes a portfolio which has a mean-reverting aggregate value. In that sense, mean-reverting portfolios are made by the investor, and cannot be simply chosen among tradable assets. A mean-reverting portfolio is characterized both by the pool of assets the investor has selected (starting with the dimension of the vector), and by the fixed nominal quantities (or weights) of each of these assets in the portfolio, which the investor also needs to set. When only two assets are considered, such baskets are usually known as long-short trading pairs. We consider in this paper baskets that are constituted by more than two assets.

\subsubsection{Mean-Reverting Baskets with Sufficient Volatility and Sparsity} A mean-reverting portfolio must exhibit sufficient mean-reversion to ensure that a contrarian strategy is profitable. To meet this requirement, investors have relied on cointegration theory~\citep{granger,maddala1998urc,johansen2005cointegration} to estimate linear combinations of assets which exhibit stationarity (and therefore mean-reversion) using historical data. We argue in this work, as we did in earlier references~\citep{alex,cuturi2013mean}, that mean-reverting strategies cannot, however, only rely on this approach to be profitable. Arbitrage opportunities can only exist if they are large enough to be traded without using too much leverage or incurring too many transaction costs. For mean-reverting baskets, this condition translates naturally into a first requirement that the gap between the basket valuation and its long term mean is large enough on average, namely that the basket price has sufficient variance or volatility. A second desirable property is that mean-reverting portfolios require trading as few assets as possible to minimize costs, namely that the weights vector of that portfolio is sparse. We propose in this work methods that maximize a proxy for mean reversion, and which can take into account at the same time constraints on variance and sparsity.
\\\,\\
We propose first in Section~\ref{s:crit} three proxies for mean reversion. Section~\ref{s:opt} defines the basket optimization problems corresponding to these quantities. We show in Section~\ref{s:sdp} that each of these problems translate naturally into semidefinite relaxations which produce either exact or approximate solutions using sparse PCA techniques. Finally, we present numerical evidence in Section~\ref{s:numres} that taking into account sparsity and volatility can significantly boost the performance of mean-reverting trading strategies in trading environments where trading costs are not negligible.

	\section{Proxies for Mean-Reversion}\label{s:crit}

	Isolating stable linear combinations of variables of multivariate time series is a fundamental problem in econometrics. A classical formulation of the problem reads as follows: given a vector valued process $x=(x_t)_t$ taking values in $\RR^n$ and indexed by time $t\in\NN$, and making no assumptions on the stationarity of each individual component of $x$, can we estimate one or many directions $y\in\RR^n$ such that the univariate process $(y^Tx_t)$ is stationary? When such a vector $y$ exists, the process $x$ is said to be cointegrated. The goal of cointegration techniques is to detect and estimate such directions $y$. Taken for granted that such techniques can efficiently isolate sparse mean reverting baskets, their financial application can be either straightforward using simple event triggers to buy, sell or simply hold the basket~\citep[\S8.6]{tsay2005analysis}, or more elaborate optimal trading strategies if one assumes that the mean-reverting basket value is a Ohrstein-Ullenbeck process, as discussed in~\citep{jurek,liu2010optimal,elie:hal-00573429}.

\subsection{Related Work and Problem Setting}
	\citet{granger} provided in their seminal work a first approach to compare two non-stationary univariate time series $(x_t,y_t)$, and test for the existence of a term $\alpha$ such that $y_t-\alpha x_t$ becomes stationary. Following this seminal work, several techniques have been proposed to generalize that idea to multivariate time series. As detailed in the survey by \citet[\S5]{maddala1998urc}, cointegration techniques differ in the modeling assumptions they require on the time series themselves. Some are designed to identify only one cointegrated relationship, whereas others are designed to detect many or all of them. Among these references, \citet{johansen} proposed a popular approach that builds upon a VAR model, as surveyed in \citep{johansen2005cointegration,johansen2009cointegration}. These approaches all discuss issues that are relevant to econometrics, such as de-trending and seasonal adjustments. Some of them focus more specifically on testing procedures designed to check whether such cointegrated relationships exist or not, rather than on the robustness of the estimation of that relationship itself. We follow in this work a simpler approach proposed by~\citet{alex}, which is to trade-off interpretability, testing and modeling assumptions for a simpler optimization framework which can be tailored to include other aspects than only stationarity.~\citet{alex} did so by adding regularizers to the predictability criterion proposed by~\citet{box1977cam}. We follow in this paper the approach we proposed in~\citep{cuturi2013mean} to design mean-reversion proxies that do not rely on any modeling assumption.
	
	Throughout this paper, we write $\symm_n$ for the $n\times n$ cone of positive definite matrices. We consider in the following a multivariate stochastic process $x=(x_t)_{t\in\NN}$ taking values in $\RR^n$. We write $\Acal_k= \Expect[x_t x_{t+k}^T], k\geq 0$ for the lag-$k$ autocovariance matrix of $x_t$ if it is finite. Using a sample path $\bx$ of $(x_t)$, where $\bx=(\bx_1,\ldots,\bx_T)$ and each $\bx_t\in\RR^n$, we write $A_k$ for the \emph{empirical} counterpart of $\Acal_k$ computed from $\bx$,
	\begin{equation}\label{eq:autos}
	A_k\defeq \frac{1}{T-k-1}\sum_{t=1}^{T-k} \tilde{\bx}_t \tilde{\bx}_{t+k}^T,\; \tilde{\bx}_t\defeq \bx_t-\frac{1}{T}\sum_{t=1}^T \bx_t.
	\end{equation}
	Given $y\in\RR^n$, we now define three measures which can all be interpreted as proxies for the mean reversion of $y^Tx_t$. \textbf{Predictability} -- defined for stationary processes by~\citet{box1977cam} and generalized for non-stationary processes by~\citet{Bewl94} -- measures how close to noise the series is. The \textbf{portmanteau} statistic~\cite{Ljun78} is used to test whether a process is white noise. Finally, the \textbf{crossing statistic}~\citep{ylvisaker1965expected} measures the probability that a process crosses its mean per unit of time. In all three cases, low values for these criteria imply a fast mean-reversion. 

	\subsection{Predictability}\label{subsec:pred}
	We briefly recall the canonical decomposition derived in~\cite{box1977cam}. Suppose that $x_t$ follows the recursion:
	\BEQ\label{eq:ar1}
	x_t= \hat{x}_{t-1} + \varepsilon_t,
	\EEQ
	where $\hat{x}_{t-1}$ is a predictor of $x_t$ built upon past values of the process recorded up to $t-1$, and $\varepsilon_t$ is a vector of i.i.d. Gaussian noise with zero mean and covariance $\Sigma \in \symm_n$ independent of all variables $(x_{r})_{r<t}$. 
	The canonical analysis in \cite{box1977cam} starts as follows. 
	\subsubsection{Univariate case} Suppose $n=1$ and thus $\Sigma\in\RR_+$, Equation (\ref{eq:ar1}) leads thus to
	\[
	\Expect[x_t^2]=\Expect[\hat{x}_{t-1}^2]+\Expect{[\varepsilon_t^2]}, \text{ thus } 1=\frac{\hat{\sigma}^2}{\sigma^2}+\frac{\Sigma}{\sigma^2},
	\] by introducing the variances $\sigma^2$ and $\hat{\sigma}^2$ of $x_t$ and $\hat{x}_t$ respectively.~\citeauthor{box1977cam} measure the \emph{predictability} of $x_t$ by the ratio
	$$
	\lambda\defeq\frac{\hat{\sigma}^2}{\sigma^2}.
	$$
	The intuition behind this variance ratio is simple: when it is small the variance of the noise dominates that of $\hat{x}_{t-1}$ and $x_t$ is dominated by the noise term; when it is large, $\hat{x}_{t-1}$ dominates the noise and $x_t$ can be accurately predicted on average.

	\subsubsection{Multivariate case} Suppose $n>1$ and consider now the univariate process $(y^Tx_t)_{t}$ with weights $y\in\RR^{n}$. Using~(\ref{eq:ar1}) we know that $y^Tx_t =y^T\hat{x}_{t-1}+y^T\varepsilon_t$, and we can measure its predicability as
	\BEQ\label{eq:pred}
	\lambda(y)\defeq \frac{y^T \hat{\Acal}_0 y}{y^T \Acal_0 y},
	\EEQ
	where $\hat{\Acal}_0$ and $\Acal_0$ are the covariance matrices of $x_t$ and $\hat{x}_{t-1}$ respectively. Minimizing predictability $\lambda(y)$ is then equivalent to finding the minimum generalized eigenvalue $\lambda$ solving
	\BEQ\label{eq:pred2}
	\det(\lambda \Acal_0 - \hat{\Acal}_0) =0.
	\EEQ
	Assuming that $\Acal_0$ is positive definite, the basket with minimum predictability will be given by $y=\Acal_0^{-1/2}y_0$, where $y_0$ is the eigenvector corresponding to the smallest eigenvalue of the matrix $\Acal_0^{-1/2} \hat{\Acal}_0 \Acal_0^{-1/2}$.

	\subsubsection{Estimation of $\lambda(y)$} All of the quantities used to define $\lambda$ above need to be estimated from sample paths. $\Acal_0$ can be estimated by $A_0$ following Equation~\eqref{eq:autos}. All other quantities depend on the predictor $\hat{x}_{t-1}$. \citeauthor{box1977cam} assume that $x_t$ follows a vector autoregressive model of order $p$ -- VAR(p) in short -- and therefore $\hat{x}_{t-1}$ takes the form,
	\[
	\hat{x}_{t-1}=\sum_{k=1}^p \Hca_k x_{t-k},
	\]
	where the $p$ matrices $(\Hca_k)$ contain each $n\times n$ autoregressive coefficients. Estimating $\Hca_k$ from the sample path $\bx$,~\citeauthor{box1977cam} solve for the optimal basket by inserting these estimates in the generalized eigenvalue problem displayed in Equation~\eqref{eq:pred2}. If one assumes that $p=1$ (the case $p>1$ can be trivially reformulated as a VAR(1) model with adequate reparameterization), then $$\hat{\Acal}_0=\Hca_1 \Acal_0 \Hca_1^T \text{ and }\Acal_1=\Acal_0 \Hca_1,$$ and thus the Yule-Walker estimator~\citep[\S3.3]{lutkepohl2005nim} of $\Hca_1$ would be $H_1=A_0^{-1} A_1$. Minimizing predictability boils down to solving in that case
	\[
	\min_{y} \hat{\lambda}(y), \; \hat{\lambda}(y)\defeq \frac{y^T \left( H_1 A_0 H_1^T\right) y}{y^T A_0 y}=\frac{y^T \left( A_1 A_0^{-1} A_1^T\right) y}{y^T A_0 y},
	\]
	which is equivalent to computing the smallest eigenvector of the matrix $A_0^{-1/2}A_1 A_0^{-1} A_1^T A_0^{-1/2}$ if the covariance matrix $A_0$ is invertible.

	The machinery of~\citeauthor{box1977cam} to quantify mean-reversion requires defining a model to form $\hat{x}_{t-1}$, the conditional expectation of $x_t$ given previous observations. We consider in the following two criteria that do without such modeling assumptions.

	\subsection{Portmanteau Criterion}\label{ss:portm}
	Recall that the {\em portmanteau} statistic of order $p$~\cite{Ljun78} of a centered univariate stationary process $x$ (with $n=1$) is given by
	$$\por_p(x)=\frac{1}{p}\sum_{i=1}^p \left(\frac{\Expect[x_t x_{t+i}]}{\Expect[x_t^2]}\right)^2$$ 
	where ${\Expect[x_t x_{t+i}]}/{\Expect[x_t^2]}$ is the $i$th order autocorrelation of $x_t$. The portmanteau statistic of a white noise process is by definition $0$ for any $p$. Given a multivariate $(n>1)$ process $x$ we write
	$$
	\phi_p(y)=\por_p(y^T x)=\frac{1}{p}\sum_{i=1}^p\left(\frac{y^T \Acal_i y}{y^T \Acal_0 y}\right)^2,
	$$
	for a coefficient vector $y\in\RR^n$. By construction, $\phi_p(y)=\phi_p(ty)$ for any $t\ne 0$ and in what follows, we will impose $\|y\|_2=1$. The quantities $\phi_p(y)$ are computed using the following estimates~\citep[p.110]{Hami94}:
	\BEQ\label{eq:portm}
	\hat{\phi}_p(y)=\frac{1}{p}\sum_{i=1}^p\left(\frac{y^T A_i y}{y^T A_0 y}\right)^2.
	\EEQ

	\subsection{Crossing Statistics}\label{ss:cross}
	\citet[\S4.1]{Kede94} define the {\em zero crossing rate} of a univariate $(n=1)$ process $x$ (its expected number of crosses around $0$ per unit of time) as
	\BEQ\label{eq:cross-rate}
	\gamma(x)=\Expect\left[\frac{\sum_{t=2}^T \one_{\{x_t x_{t-1}\leq 0\}}}{T-1}\right],
	\EEQ
	A result known as the cosine formula states that if $x_t$ is an autoregressive process of order one AR(1), namely if $|a|<1$, $\varepsilon_t$ is i.i.d. standard Gaussian noise and $x_t=a x_{t-1} + \varepsilon_t$, then~\citep[\S4.2.2]{Kede94}:
	\[
	\gamma(x)=\frac{\arccos(a)}{\pi}.
	\]
	Hence, for AR(1) processes, minimizing the first order autocorrelation $a$ also directly maximizes the crossing rate of the process $x$. For $n>1$, since the first order autocorrelation of $y^Tx_t$ is equal to $y^T\Acal_1y$, we propose to minimize $y^T\Acal_1y$ and ensure that all other absolute autocorrelations $\abs{y^T\Acal_ky}$, $k>1$ are small.

	\section{Optimal Baskets}\label{s:opt}
Given a centered multivariate process $\bx$, we form its covariance matrix $A_0$ and its $p$ autocovariances $(A_1,\ldots,A_p)$. Because $y^TAy=y^T(A+A^T)y/2$, we symmetrize all autocovariance matrices $A_i$. We investigate in this section the problem of estimating baskets that have maximal mean reversion (as measured by the proxies proposed in Section\ref{s:crit}), while being at the same time sufficiently volatile and supported by as few assets as possible. The latter will be achieved by selecting portfolios $y$ that have a small ``0-norm'', namely that the number of non-zero components in $y$, 
$$\|y\|_0\defeq \#\{1\leq i\leq d | y_i\ne 0\},$$ is small. The former will be achieved by selecting portfolios whose aggregated value exhibits a variance over time that exceeds a given threshold $\nu>0$. Note that for the variance of $(y^Tx_t)$ to exceed a level $\nu$, the largest eigenvalue of $A_0$ must necessarily be larger than~$\nu$, which we always assume in what follows. Combining these two constraints, we propose three different mathematical programs that reflect these trade-offs.

\subsection{Minimizing Predictability}\label{ss:opt-pred}
	Minimizing Box-Tiao's predictability $\hat{\lambda}$ defined in \S\ref{subsec:pred} while ensuring that both the variance of the resulting process exceeds $\nu$ and that the vector of loadings is sparse with a 0-norm equal to $k$, means solving the following program:
	\BEQ\label{eq:P1}\tag{P1}
	\BA{ll}
	\mbox{minimize} & y^T M y\\
	\mbox{subject to} & y^T A_0y\geq \nu,\\
	& \|y\|_2=1,\\
	& \|y\|_0=k,
	\EA\EEQ
	in the variable $y\in\RR^n$ with $M\defeq A_1 A_0^{-1} A_1^T$, where $M,A_0\in\symm_n$. Without the normalization constraint $\|y\|_2=1$ and the sparsity constraint $\|y\|_0=k$, problem~\eqref{eq:P1} is equivalent to a generalized eigenvalue problem in the pair $(M,A_0)$. That problem quickly becomes unstable when $A_0$ is ill-conditioned or $M$ is singular. Adding the normalization constraint $\|y\|_2=1$ solves these numerical problems.

	\subsection{Minimizing the Portmanteau Statistic}\label{ss:opt-portm}
	Using a similar formulation, we can also minimize the order $p$ portmanteau statistic defined in \S\ref{ss:portm} while ensuring a minimal variance level $\nu$ by solving:
	\BEQ\BA{ll}\label{eq:P2}\tag{P2}
	\mbox{minimize} &\sum_{i=1}^{p}\left(y^T A_i y\right)^2\\
	\mbox{subject to} &  y^T A_0y \geq \nu,\\ 
	& \|y\|_2=1,\\
	& \|y\|_0=k,
	\EA\EEQ
	in the variable $y\in\RR^n$, for some parameter $\nu>0$. Problem~\eqref{eq:P2} has a natural interpretation: the objective function directly minimizes the portmanteau statistic, while the constraints normalize the norm of the basket weights to one, impose a variance larger than~$\nu$ and impose a sparsity constraint on $y$. 

	\subsection{Minimizing the Crossing Statistic}\label{ss:opt-portm2}
	Following the results in \S\ref{ss:cross}, maximizing the crossing rate while keeping the rest of the autocorrelogram low,
	\BEQ\BA{ll}\label{eq:P3}\tag{P3}
	\mbox{minimize} & y^TA_1y + \mu \sum_{k=2}^{p}\left(y^T A_k y\right)^2\\
	\mbox{subject to} &  y^T A_0y \geq \nu,\\ 
	& \|y\|_2=1,\\	
	& \|y\|_0=k,
	\EA\EEQ
	in the variable $y\in\RR^n$, for some parameters $\mu,\nu>0$, will produce processes that are close to being AR(1), while having a high crossing rate.

	\section{Semidefinite Relaxations and Sparse Components}\label{s:sdp}
	Problems~\eqref{eq:P1},~\eqref{eq:P2} and ~\eqref{eq:P3} are not convex, and can be in practice extremely difficult to solve, since they involve a sparse selection of variables. We detail in this section convex relaxations to these problems which can be used to derive relevant sub-optimal solutions.
	
	\subsection{A Semidefinite Programming Approach to Basket Estimation}\label{subsec:asemidefinite}
	We propose to relax problems~\eqref{eq:P1},~\eqref{eq:P2} and ~\eqref{eq:P3} into Semidefinite Programs (SDP) \citep{vandenberghe1996semidefinite}. We show that these semidefinite programs can handle naturally sparsity and volatility constraints while still aiming at mean-reversion. In some restricted cases, one can show that these relaxations are tight, in the sense that they solve exactly the programs described above. In such cases, the true solution $y^\star$ of some of the programs above can be recovered using their corresponding SDP solution $Y^\star$. 
	
	However, in most of the cases we will be interested in, such a correspondence is not guaranteed and these SDP relaxations can only serve as a guide to propose solutions to these hard non-convex problems when considered with respect to vector $y$. To do so, the optimal solution $Y^\star$ needs to be \emph{deflated} from a large rank $d\times d$ matrix to a rank one matrix $yy^T$, where $y$ can be considered a good candidate for basket weights. A typical approach to deflate a positive definite matrix into a vector is to consider its eigenvector with the leading eigenvalue. Having sparsity constraints in mind, we propose to apply a heuristic grounded on sparse-PCA~\citep{zou2006sparse,d2007direct}. Instead of considering the lead eigenvector, we recover the leading \emph{sparse} eigenvector of $Y^\star$ (with a $0$-norm constrained to be equal to $k$). Several efficient algorithmic approaches have been proposed to solve approximately that problem; we use the SPASM toolbox \citep{sjostrand2012spasm} in our experiments.

	\subsection{Predictability}
	We can form a convex relaxation of the predictability optimization problem~\eqref{eq:P1} over the variable $y\in\RR^n$,
	\[\BA{ll}
	\mbox{minimize} & y^T M y\\
	\mbox{subject to} & y^T A_0y\geq \nu\\
	& \|y\|_2=1,\\
	& \|y\|_0=k,
	\EA\]
	by using the lifting argument of~\citet{Lova91}, \ie\, writing $Y=yy^T$, to solve now the problem using a semidefinite variable $Y$, and by introducing a sparsity-inducing regularizer on $Y$ which considers the $L_1$ norm of $Y$,
	$$
	\norm{Y}_1\defeq \sum_{ij}\abs{Y_{ij}},
	$$
	 so that Problem~\eqref{eq:P1} becomes (here $\rho>0$),
	\[\BA{ll}
	\mbox{minimize} & \Tr(MY) + \rho \norm{Y}_1\\
	\mbox{subject to} & \Tr(A_0Y)\geq\nu\\
	& \Tr(Y)=1,~\Rank(Y)=1,~Y\succeq 0.
	\EA\]
	We relax this last problem further by dropping the rank constraint, to get
	\BEQ\BA{ll}\label{eq:SDP1}\tag{SDP1}
	\mbox{minimize} & \Tr(MY)  + \rho \norm{Y}_1\\
	\mbox{subject to} & \Tr(A_0Y)\geq\nu\\
	& \Tr(Y)=1,~Y\succeq 0
	\EA\EEQ
	which is a convex semidefinite program in $Y\in\symm_n$.

	\subsection{Portmanteau}
	Using the same lifting argument and writing $Y=yy^T$, we can relax problem~\eqref{eq:P2} by solving
	\BEQ\BA{ll}\label{eq:SDP2}\tag{SDP2}
	\mbox{minimize} & \sum_{i=1}^p \Tr(A_iY)^2 + \rho \norm{Y}_1\\
	\mbox{subject to} & \Tr(BY)\geq\nu\\
	& \Tr(Y)=1,~Y\succeq 0,
	\EA\EEQ
	a semidefinite program in $Y\in\symm_n$.
	
	\subsection{Crossing Stats}
	As above, we can write a semidefinite relaxation for problem~\eqref{eq:P3}:
	\BEQ\BA{ll}\label{eq:SDP3}\tag{SDP3}
	\mbox{minimize} & \Tr(A_1Y)+ \mu \sum_{i=2}^p \Tr(A_iY)^2 + \rho \norm{Y}_1\\
	\mbox{subject to} & \Tr(BY)\geq\nu\\
	& \Tr(Y)=1,~Y\succeq 0
	\EA\EEQ	

	\subsubsection{Tightness of the SDP Relaxation in the Absence of Sparsity Constraints} Note that for the crossing stats criterion (with $p=1$ and no quadratic term in $Y$) criteria, the original problem \ref{eq:P3} and its relaxation \ref{eq:SDP3} are equivalent, taken for granted that no sparsity constraint is considered in the original problems and $\mu$ set to $0$ in the relaxations. This relaxations boil down to an SDP's that only has a linear objective, a linear constraint and a constraint on the trace of $Y$. In that case, \citet{Bric61} showed that the range of two quadratic forms over the unit sphere is a convex set when the ambient dimension $n\geq 3$, which means in particular that for any two square matrices $A,B$ of dimension $n$
	\BEAS
	&\left\{(y^TAy,y^TBy): y\in\RR^n,~\|y\|_2=1\right\}=&\\
	&\left\{(\Tr(AY),\Tr(BY)): Y\in\symm_n,~\Tr Y=1,~Y\succeq 0\right\}&
	\EEAS
	We refer the reader to~\citep[\S II.13]{Barv02} for a more complete discussion of this result. As remarked in~\citep{cuturi2013mean}, the same equivalence holds for \ref{eq:P1} and \ref{eq:SDP1}. This means that, in the case where $\rho,\mu=0$ and the 0-norm of $y$ is \emph{not} constrained, for any solution $Y^\star$ of the relaxation~\eqref{eq:SDP1} there exists a vector $y^\star$ which satisfies $\norm{y}_2^2=\Tr(Y^\star)=1$, $y^{\star T} A_0 y^\star=\Tr(BY^\star)$ and $y^{\star T}My^\star=\Tr(MY^\star)$ which means that $y^\star$ is an optimal solution of the original problem~\eqref{eq:P1}. \citet[App.\,B]{Boyd:1072} show how to explicitly extract such a solution~$y^\star$ from a matrix $Y^\star$ solving~\eqref{eq:SDP1}. This result is however mostly anecdotical in the context of this paper, in which we look for sparse and volatile baskets: using these two regularizers breaks the tightness result between the original problems in $\RR^d$ and their SDP counterparts.


	\section{Numerical Experiments}\label{s:numres}

	\begin{figure}
	\centering
 	\includegraphics[width=.7\textwidth]{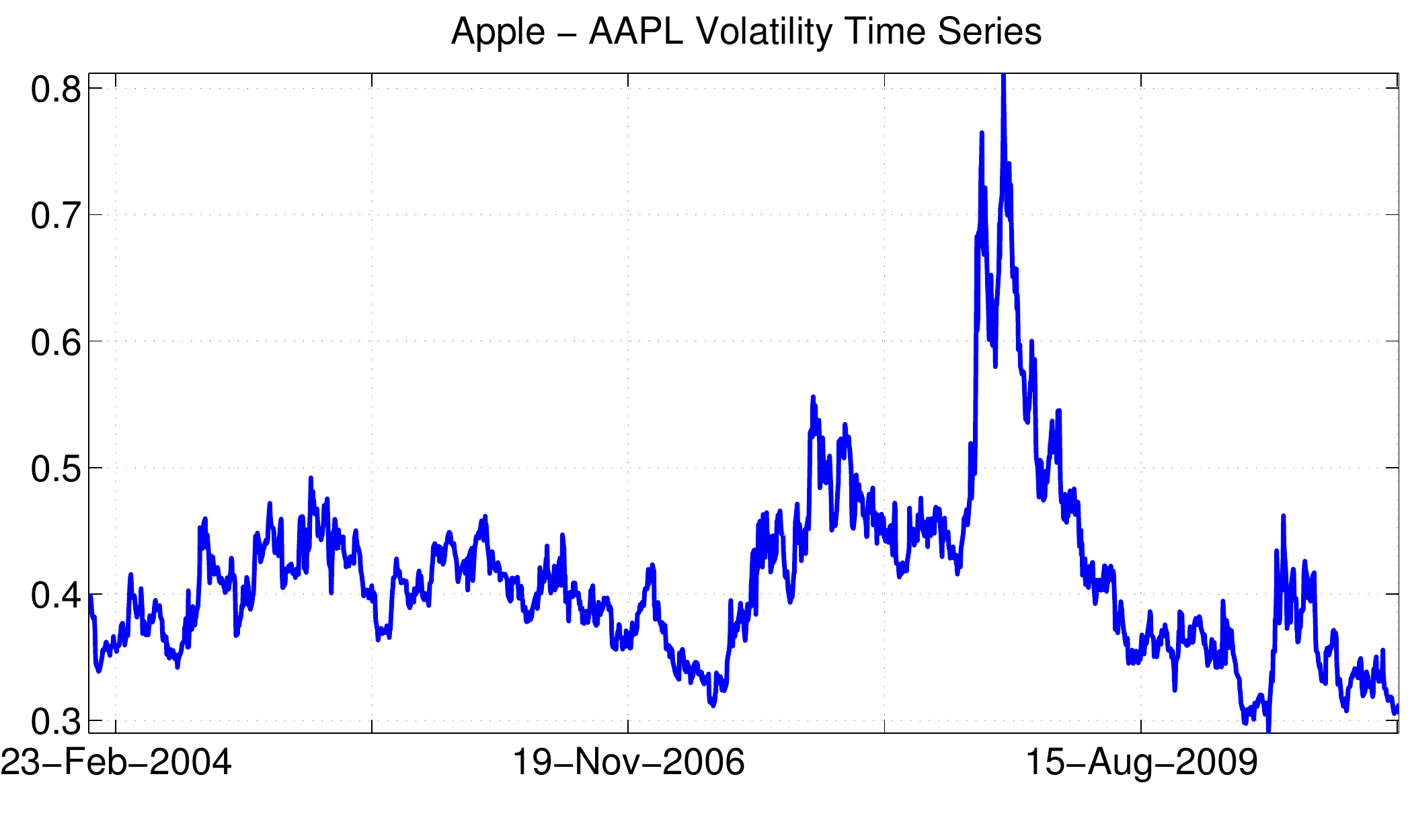}
	\caption{\textbf{Option implied volatility} for Apple between January 4 2004 and December 30 2010.}\label{fig:vol}
	\end{figure}

	In this section, we evaluate the ability of our techniques to extract mean-reverting baskets with sufficient variance and small 0-norm from a universe of tradable assets. We measure performance by applying to these baskets a trading strategy designed specifically for mean-reverting processes. We show that, under realistic trading costs assumptions, selecting sparse and volatile mean-reverting baskets translates into lower incurred costs and thus improves the performance of trading strategies.

	\subsection{Historical Data} We consider daily time series of option implied volatilities for 210 stocks from January 4 2004 to December 30 2010. A key advantage of using option implied volatility data is that these numbers vary in a somewhat limited range. Volatility also tends to exhibit regime switching, hence can be considered piecewise stationary, which helps in extracting structural relationships. We plot a sample time series from this dataset in Figure~\ref{fig:vol} that corresponds to the implicit volatility of Apple's stock. In what follows, we mean by asset the implied volatility of any of these stocks, whose value can be efficiently replicated using option portfolios.

	\subsection{Mean-reverting Basket Estimators} We compare the three basket selection techniques detailed here, \textbf{predictability}, \textbf{portmanteau} and \textbf{crossing statistic}, implemented with varying targets for both sparsity and volatility, with two cointegration estimators that build upon principal component analysis \citep[\S5.5.4]{maddala1998urc}. By the label `PCA' we mean in what follows the eigenvector with smallest eigenvalue of the covariance matrix $A_0$ of the process \citep{stock1988tct}. By `sPCA' we mean the sparse eigenvector of $A_0$ with 0-norm $k$ that has the smallest eigenvalue, which can be simply estimated by computing the leading sparse eigenvector of $\lambda I-A_0$ where $\lambda$ is bigger than the leading eigenvalue of $A_0$. This sparse principal component of the covariance matrix $A_0$ should not be confused with our utilization of sparse PCA in Section~\ref{subsec:asemidefinite} as a way to recover a vector solution from the solution of a positive semidefinite problem. Note also that techniques based on principal components do not take explicitly variance levels into account when estimating the weights of a co-integrated relationship.

	\subsection{\citeauthor{jurek} Trading Strategy}
\begin{figure}
		
	\hskip-2.5cm\includegraphics[width=1.4\textwidth]{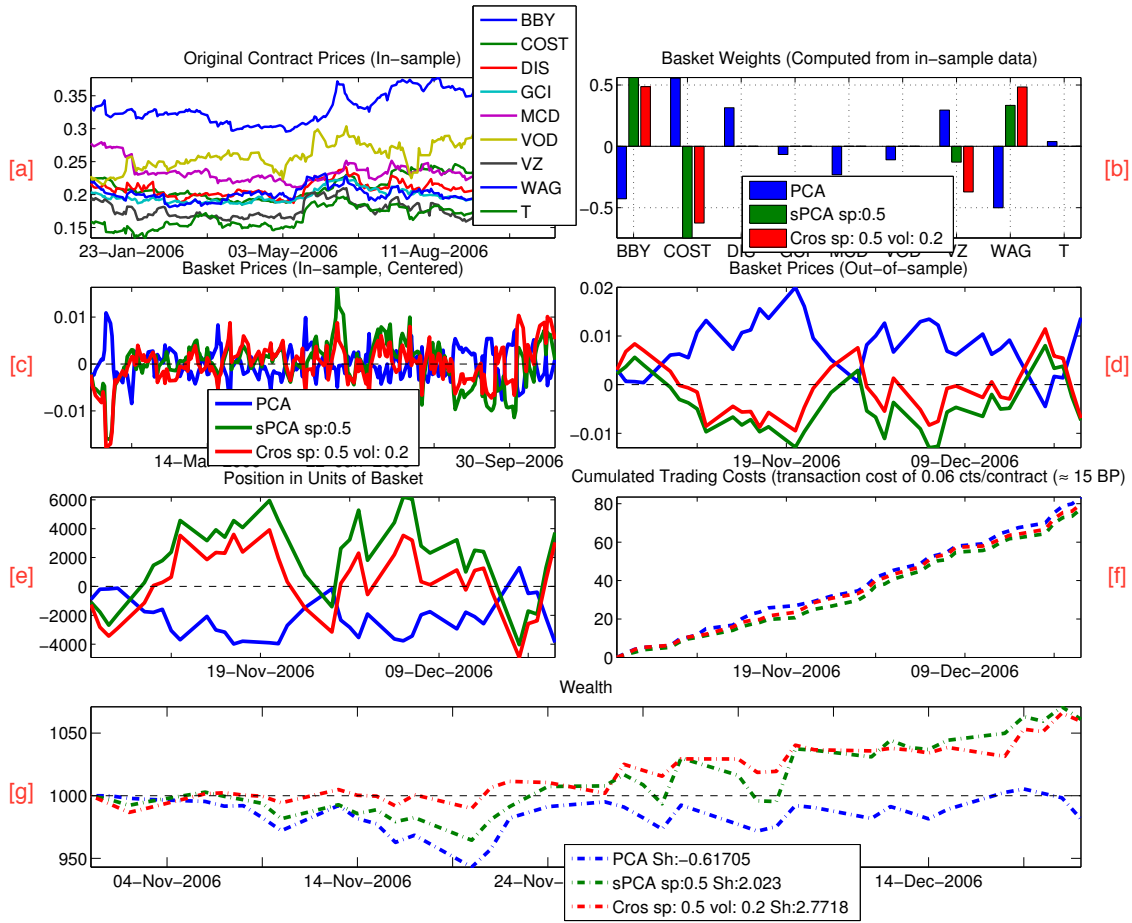}
\caption{\textbf{Three sample trading experiments, using the PCA, sparse PCA and the Crossing Statistics estimators}. [a] Pool of 9 volatility time-series selected using our fast PCA selection procedure. [b] Basket weights estimated with in-sample data using either the eigenvector of the covariance matrix with smallest eigenvalue, the smallest eigenvector with a sparsity constraint of $k=\lfloor 0.5 \times 9\rfloor=4$ and the Crossing Statistics estimator with a volatility threshold of $\nu=0.2$, \ie a constraint on the basket's variance to be larger than $0.2 \times$ the median variance of all $8$ assets. [c] Using these 3 procedures, the time series of the resulting basket price in the in-sample part [c] and out-sample parts [d] are displayed.  [e] Using the \cite{jurek} trading strategy results in varying positions (expressed as units of baskets) during the out-sample testing phase. [f] Transaction costs that result from trading the assets to achieve such positions accumulate over time. [g] Taking both trading gains and transaction costs into account, the net wealth of the investor for each strategy can be computed (the Sharpe over the test period is displayed in the legend). Note how both sparsity and volatility constraints translate into portfolios composed of less assets, but with a higher variance.}\label{fig:syn}
	\end{figure}

	While option implied volatility is not directly tradable, it can be synthesized using baskets of call options, and we assimilate it to a tradable asset with (significant) transaction costs in what follows. For baskets of volatilities isolated by the techniques listed above, we apply the~\citep{jurek} strategy for log utilities to the basket process recording out of sample performance.~\citeauthor{jurek} propose to trade a stationary autoregressive process $(x_t)_{t}$ of order $1$ and mean $\mu$ governed by the equation $x_{t+1} = \rho x_t +\sigma \varepsilon_t$,
	where $\abs{\rho}<1$, by taking a position $N_t$ in the asset $x_t$ which is proportional to
	\begin{equation}\label{eq:jurek}
	N_t = \frac{\rho (\mu-x_t)}{\sigma^2}W_t
	\end{equation}
	In effect, the strategy advocates taking a long (resp. short) position in the asset whenever it is below (resp. above) its long-term mean, and adjust the position size to account for the volatility of $x_t$ and its mean reversion speed $\rho$. Given basket weights $y$, we apply standard AR estimation procedures on the in-sample portion of $y^T\bx$ to recover estimates for $\hat{\rho}$ and $\hat{\sigma}$ and plug them directly in Equation~\eqref{eq:jurek}. This approach is illustrated for two baskets in Figure~\ref{fig:syn}.

	\subsection{Transaction Costs}
	We assume that fixed transaction costs are negligible, but that transaction costs per contract unit are incurred at each trading date. We vary the size of these costs across experiments to show the robustness of the approaches tested here to trading costs fluctuations. We let the transaction cost per contract unit vary between 0.03 and 0.17 cents by increments of 0.02 cents. Since the average value of a contract over our dataset is about 40 cents, this is akin to considering trading costs ranging from about 7 to about 40 Base Points (BP), that is 0.07 to 0.4\%. 

	\subsection{Experimental Setup}
	We consider 20 sliding windows of one year (255 trading days) taken in the history, and consider each of these windows independently.  Each window is split between 85\% of days to estimate and 15\% of days to test-trade our models, resulting in 38 test-trading days. We do not recompute the weights of the baskets during the test phase. The 210 stock volatilities (assets) we consider are grouped into 13 subgroups, depending on the economic sector of their stock. This results in 13 sector pools whose size varies between 3 assets and 43 assets. We look for mean-reverting baskets in each of these 13 sector pools.
	
	Because all combinations of stocks in each of the 13 sector pools may not necessarily mean-reverting, we select smaller candidate pools of $n$ assets through a greedy backward-forward minimization scheme, where $8\leq n\leq 12$. To do so, we start with an exhaustive search of all pools of size 3 within the sector pool, and proceed by adding or removing an asset using the PCA estimator (the smallest eigenvalue of the covariance matrix of a set of assets). We use the PCA estimator in that backward-forward search because it is the fastest to compute. We score each pool using that PCA statistic, the smaller meaning the better. We generate up to 200 candidate pools per each of the 13 sector pools. Out of all these candidate pools, we keep the best 50 in each window, and use then our cointegration estimation approaches separately on these candidates. One such pool was, for instance, composed of the stocks \texttt{\{BBY,COST,DIS,GCI,MCD,VOD,VZ,WAG,T\}} observed during the year 2006. Figure~\ref{fig:syn} provides a closeup on that universe of stocks, and shows the results of three trading experiments using either PCA, sparse PCA or the Crossing Stats estimator to build trading strategies.

	\subsection{Results}
\subsubsection{Robustness of Sharpe Ratios to Costs}	In Figure~\ref{fig:sharpe}, we plot the average of the Sharpe ratio over the $922$ baskets estimated in our experimental set versus transaction costs. We consider different PCA settings as well as our three estimators using, in all 3 cases, the variance bound $\nu$ to be $0.3$ times the median of all variances of assets available in a given asset pool, and the 0-norm to be equal to 0.3 times the size of the universe (itself between 8 and 12). We observe that Sharpe ratios decrease the fastest for the naive PCA based method, this decrease being somewhat mitigated when adding a constraint on the 0-norm of the basket weights obtained with sparse PCA. Our methods require, in addition to sparsity, enough volatily to secure sufficient gains. These empirical observations agree with the intuition of this paper: simple cointegration techniques can produce synthetic baskets with high mean-reversion, large support, low variance. Trading a portfolio with low variance which is supported by multiple assets translates in practice into high trading costs which can damage the overall performance of the strategy. Both sparse PCA and our techniques manage instead to achieve a trade-off between desirable mean-reversion properties and, at the same time, control for sufficient variance and small basket size to allow for lower overall transaction costs.
	
\subsubsection{Tradeoffs Between Mean Reversion, Sparsity, and Volatility }	
\begin{figure}[ht]
	\centering
	\includegraphics[width=.8\textwidth]{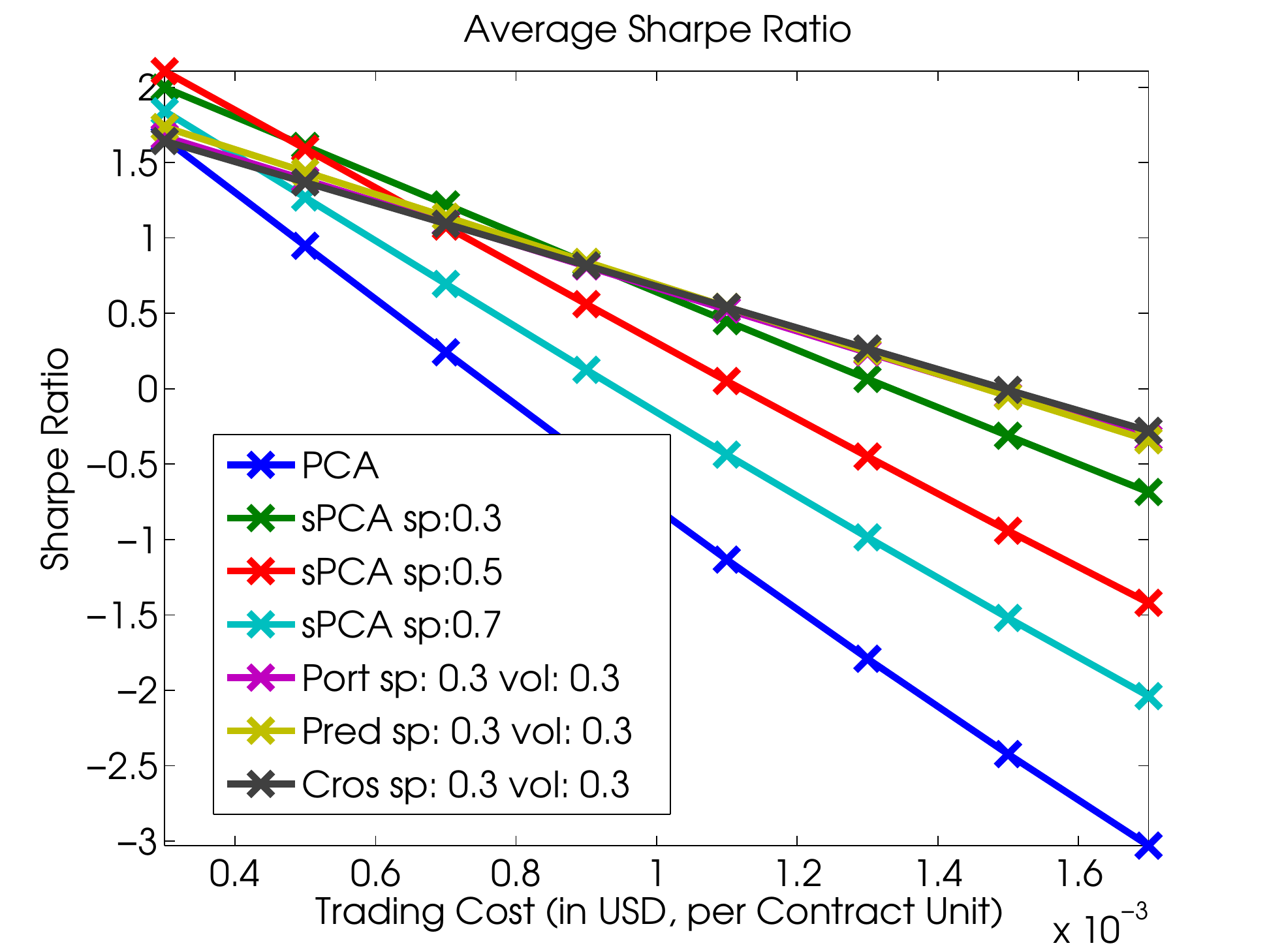}
	\caption{Average Sharpe ratio for the~\citet{jurek} trading strategy captured over about 922 trading episodes, using different basket estimation approaches. These 922 trading episodes were obtained by considering 7 disjoint time-windows in our market sample, each of a length of about one year. Each time-window was divided into 85\% in-sample data to estimate baskets, and 15\% outsample to test strategies. On each time-window , the set of 210 tradable assets during that period was clustered using sectorial information, and each cluster screened (in the in-sample part of the time-window) to look for the most promising baskets of size between 8 and 12 in terms of mean reversion, by choosing greedily subsets of stocks that exhibited the smallest minimal eigenvalues in their covariance matrices. For each trading episode, the same universe of stocks was fed to different mean-reversion algorithms. Because volatility time-series are bounded and quite stationary, we consider the PCA approach, which uses the eigenvector with the smallest eigenvalue of the covariance matrix of the time-series to define a cointegrated relationship. Besides standard PCA, we have also consider sparse PCA eigenvectors with minimal eigenvalue, with the size $k$ of the support  of the eigenvector (the size of the resulting basket) constrained to be 30\%, 50\% or 70\% of the total number of considered assets. We consider also the portmanteau, predictability and crossing stats estimation techniques with variance thresholds of $\nu=0.2$ and a support whose size $k$ (the number of assets effectively traded) is targeted to be about $30\%$ of the size of the considered universe (itself between 8 and 12). As can be seen in the figure, the sharpe ratios of all trading approaches decrease with an increase in transaction costs. One expects sparse baskets to perform better under the assumption that costs are high, and this is indeed observed here. Because the relationship between sharpe ratios and transaction costs can be efficiently summarized as being a linear one, we propose in the plots displayed in Figures~\ref{fig:sharpeCrossing} and ~\ref{fig:sharpeCrossing2} a way to summarize the lines above with two numbers each: their intercept (Sharpe level in the quasi-absence of costs) and slope (degradation of Sharpe as costs increase). This visualization is useful to observe how sparsity (basket size) and volatility thresholds influence the robustness to costs of the strategies we propose. This visualization allows us to observe how performance is influenced by these parameter settings.\label{fig:sharpe}}
	\end{figure}

\begin{figure}[ht]
		\centering
		\includegraphics[width=\textwidth]{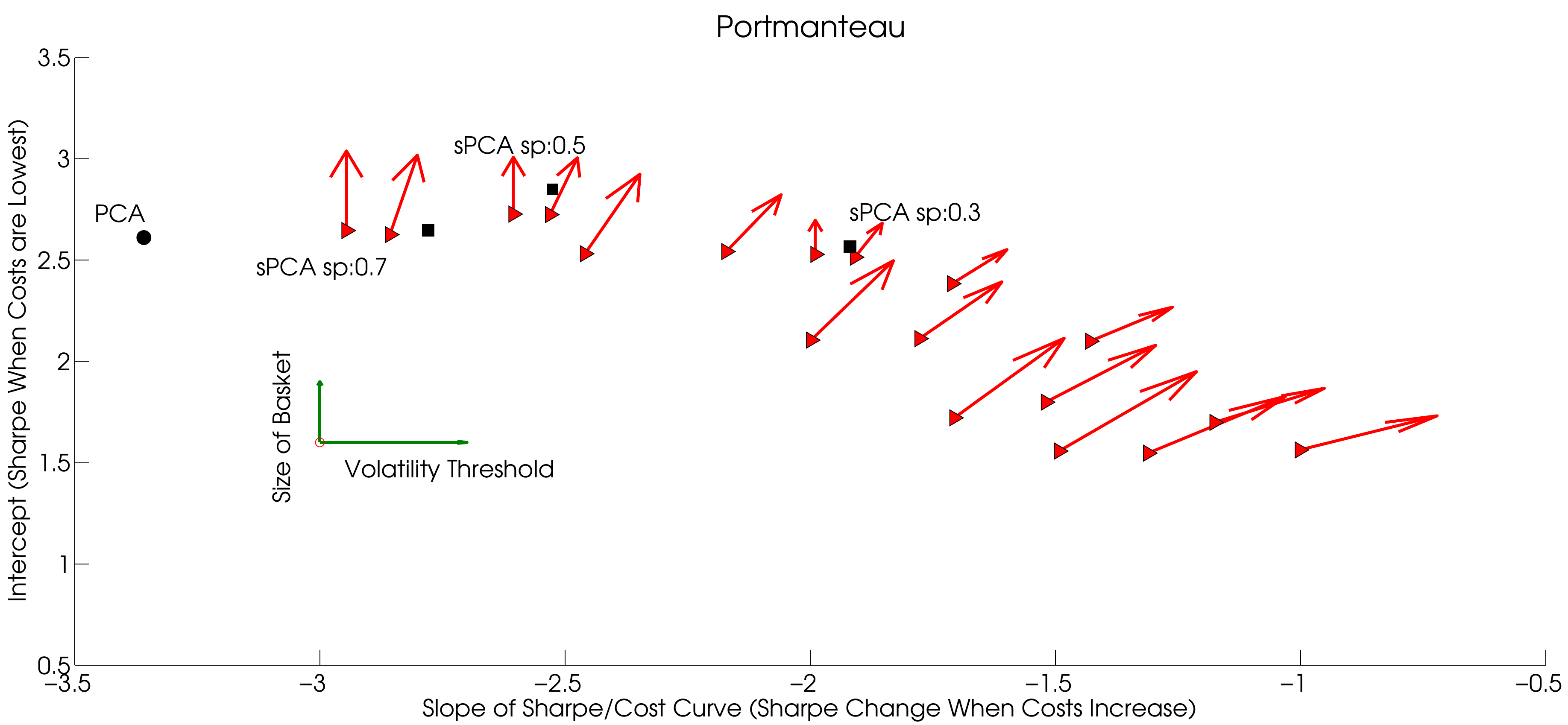}\\
		(a)
		\includegraphics[width=\textwidth]{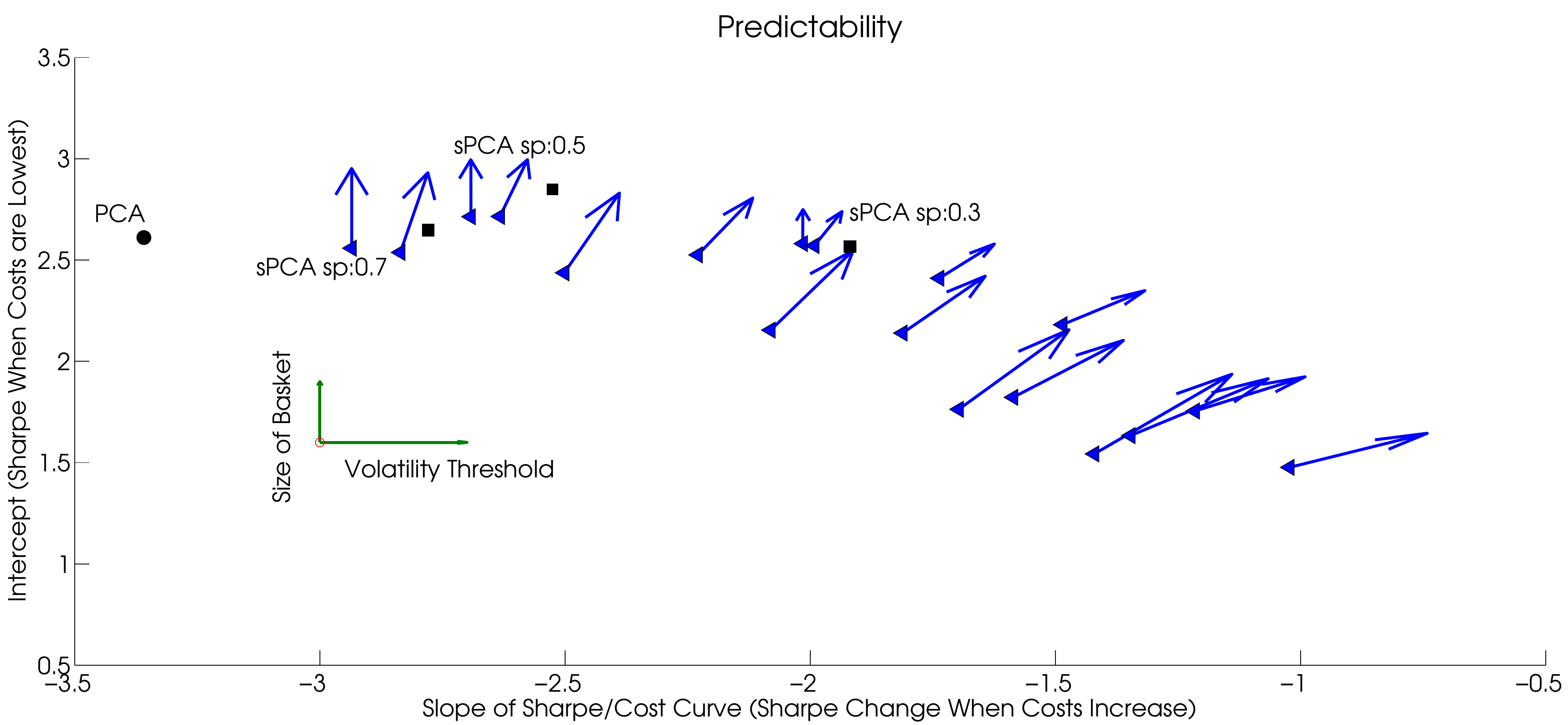}\\
		(b)
	\caption{Relationships between Sharpe in a low cost setting (intercept) in the $x$-axis and robustness of Sharpe to costs (slope of Sharpe/costs curve) of a different estimators implemented with varying volatility levels $\nu$ and sparsity levels $k$ parameterized as a multiple of the universe size. Each colored square in the figures above corresponds to the performance of a given estimator (Portmanteau in subfigure $(a)$, Predictability in subfigure $(b)$) using different parameters for $\nu\in\{0,0.1,0.2,0.3,0.4,0.5\}$ and $u\in\{0.3,0.5,0.7\}$. The parameters used for each experiment are displayed using an arrow whose vertical length is proportional to $\nu$ and horizontal length is proportional to $u$.\label{fig:sharpeCrossing}}
	\end{figure}

\begin{figure}[ht]
		\centering
		\includegraphics[width=\textwidth]{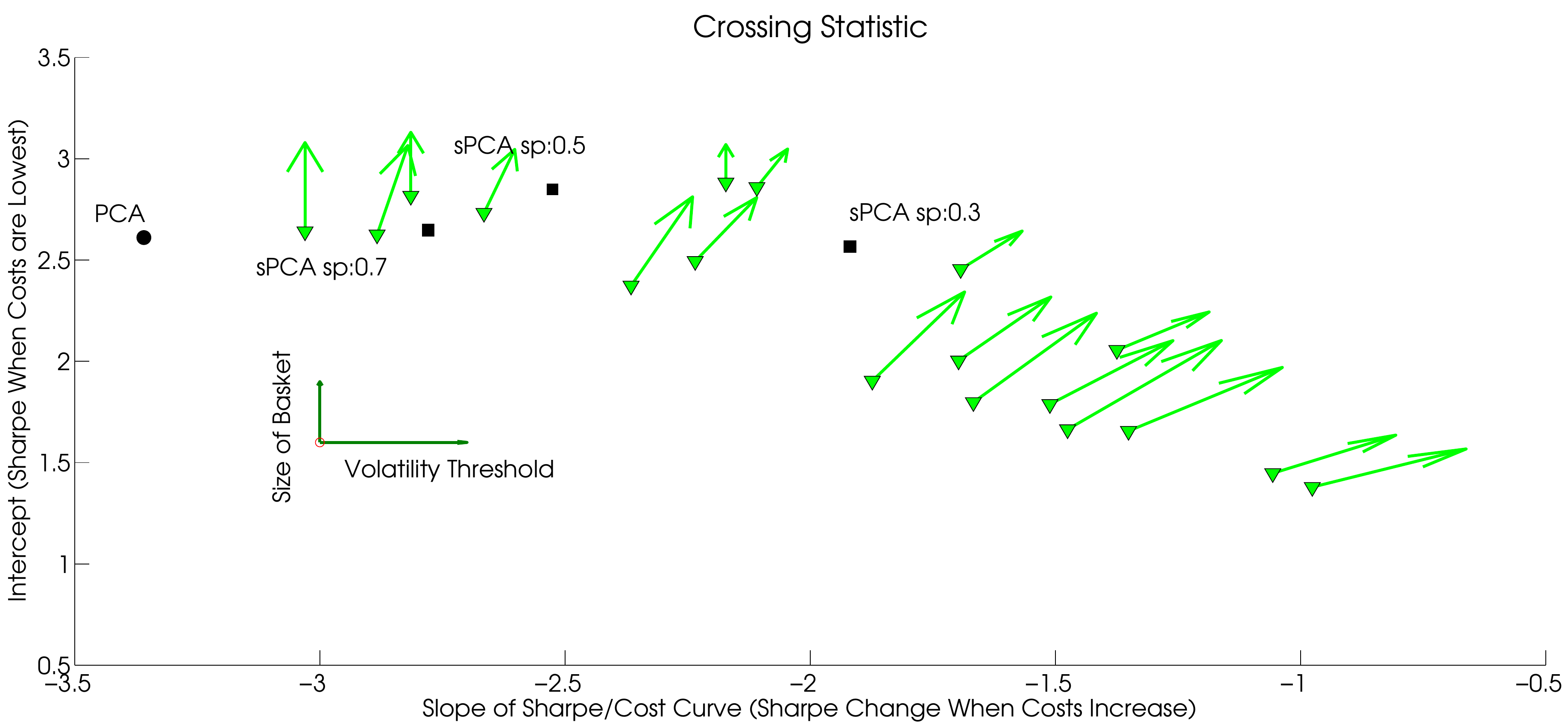}\\
		(c)
	\caption{Same setting as Figure~\ref{fig:sharpeCrossing}, using the crossing statistics (c).\label{fig:sharpeCrossing2}}
\end{figure}

In the plots of Figure~\ref{fig:sharpeCrossing} and~\ref{fig:sharpeCrossing2}, this analysis is further detailed by considering various settings for $\nu$ (volatility threshold) and $k$. To improve the legibility of these results we summarize, following the observation in Figure~\ref{fig:sharpe} that the relationship between Sharpes and transactions costs seems almost linear, each of these curves by two numbers: an intercept level (Sharpe ratio when costs are low) and a slope (degradtion of Sharpe as costs increase). Using these two numbers, we locate all considered strategies in the intercept/slope plane. We first show the spectral techniques, PCA and sPCA with different levels of sparsity, meaning that $k$ is set to $\lfloor u \times d\rfloor$ where $u\in\{0.3,0.5,0.7\}$ and $d$ is the size of the original basket. Each of the three estimators we propose is studied in a separate plot. For each we present various results characterized by two numbers: a volatility threshold $\nu\in\{0,0.1,0.2,0.3,0.4,0.5\}$ and a sparsity level $u\in\{0.3,0.5,0.7\}$. To avoid cumbersome labels, we attach an arrow to each point: the arrow's length in the vertical direction is equal to $u$ and characterizes the size of the basket, the horizontal length is equal to $\nu$ and characterizes the volatility level. As can be seen in these 3 plots, an interesting interplay between these two factors allows for a continuum of strategies that trade mean-reversion (and thus Sharpe levels) for robustness to cost level.

%
%
%
%

	%
	%
	%
	%

	\section{Conclusion}
	We have described three different criteria to quantify the amount of mean reversion in a time series. For each of these criteria, we have detailed a tractable algorithm to isolate a vector of weights that has optimal mean reversion, while constraining both the variance (or signal strength) of the resulting univariate series to be above a certain level and its 0-norm to be at a certain level. We show that these bounds on variance and support size, together with our new criteria for mean reversion, can significantly improve the performance of mean reversion statistical arbitrage strategies and provide useful controls to adjust mean-reverting strategies to varying trading conditions, notably liquidity risk and cost environment.

\bibliographystyle{plainnat}

\end{document}